\documentclass[conference]{IEEEtran}
\IEEEoverridecommandlockouts

\usepackage{cite}
\usepackage{amsmath,amssymb,amsfonts}
\usepackage{algorithmic}
\usepackage{graphicx}
\usepackage{textcomp}
\usepackage{xcolor}
\usepackage{array}
\def\BibTeX{{\rm B\kern-.05em{\sc i\kern-.025em b}\kern-.08em
    T\kern-.1667em\lower.7ex\hbox{E}\kern-.125emX}}

\begin{document}

\title{RIS-Enabled Energy-Efficient ISAC for Vehicular Applications}
\vspace{-0.5cm}
\author{\IEEEauthorblockN{Shengyu~Yang$^{\ast}$, Yuhan~Wang$^{\dag}$, Shuhao~Zeng$^{\dag}$$^{\S}$, Haobo~Zhang$^{\dag}$$^{\ddagger}$, Boya~Di$^{\ast}$, Lingyang~Song$^{\dag}$$^{\ast}$}
\IEEEauthorblockA{$^{\ast}$School of Electronics, Peking University, Beijing, China.\\
$^{\dag}$School of Electronic and Computer Engineering, Peking University Shenzhen Graduate School, China.\\
$^{\S}$Department of Electrical and Computer Engineering, Princeton University, USA.\\
$^{\ddagger}$Department of Engineering, University of Cambridge, Cambridge, UK.}
Email: \{yang\_sy, yuhan.wang\}@stu.pku.edu.cn, \{shuhao.zeng, haobo.zhang, diboya, lingyang.song\}@pku.edu.cn
}

\maketitle

\begin{abstract}

By incorporating integrated sensing and communication (ISAC) into vehicle-to-infrastructure (V2I) networks, roadside units~(RSUs) can support data transmission while providing additional sensing capabilities, thereby enabling intelligent transportation services. By deploying large-scale antenna arrays at RSU, the V2I network can realize more reliable connectivity and more accurate vehicle tracking by harnessing the significant beamforming gains provided by the enlarged antenna aperture. However, it is not energy efficient to realize such arrays using conventional phased arrays, which rely on numerous power-hungry phase shifters. To address this, this demo presents the first reconfigurable intelligent surface (RIS)-based ISAC-empowered vehicular network prototype, which operates at sub-6 GHz band to be compatible with existing vehicular systems. The RIS has a low power consumption of $6.8$~W. Experimental results show that compared with the system without RIS, the proposed RIS-based prototype enables more accurate vehicle trajectory tracking with an average localization error of $0.11$~m and supports more robust data transmission, as evidenced by a $41.9\%$ reduction in error vector magnitude~(EVM). These results validate the effectiveness of the RIS-based ISAC system for supporting vehicular networks.

\end{abstract}


\vspace{-0.2cm}
\section{Demo Description}

\subsection{Background}

Future vehicular networks are expected to support reliable and low-latency data transmission while providing real-time sensing capabilities for intelligent transportation services. By incorporating integrated sensing and communication (ISAC)~\cite{zeng2026} into vehicle-to-infrastructure (V2I) networks, roadside units (RSUs) can provide communication services to vehicles while tracking their trajectories, thereby supporting traffic monitoring and driving assistance. To meet these requirements, roadside infrastructure needs a large-aperture antenna array with flexible beamforming capability, so as to achieve higher beam gain, maintain reliable V2I links, and improve the accuracy of vehicle localization and tracking. However, conventional phased arrays require a large number of phase shifters, leading to high hardware cost and power consumption~\cite{dai2020ris,Zeng2024}. These limitations make the large-scale deployment of roadside antenna arrays challenging.

To address this issue, we propose to employ a reconfigurable intelligent surface (RIS) to realize a roadside ISAC system. By altering its large number of low-power programmable elements, the RIS enables flexible beamforming. Therefore, it provides a low-power and low-cost implementation with great potential for large-scale antenna array deployment, while enhancing both communication and sensing performance for V2I service.

\begin{figure}[!t]
	\centering
	\includegraphics[width=0.48\textwidth]{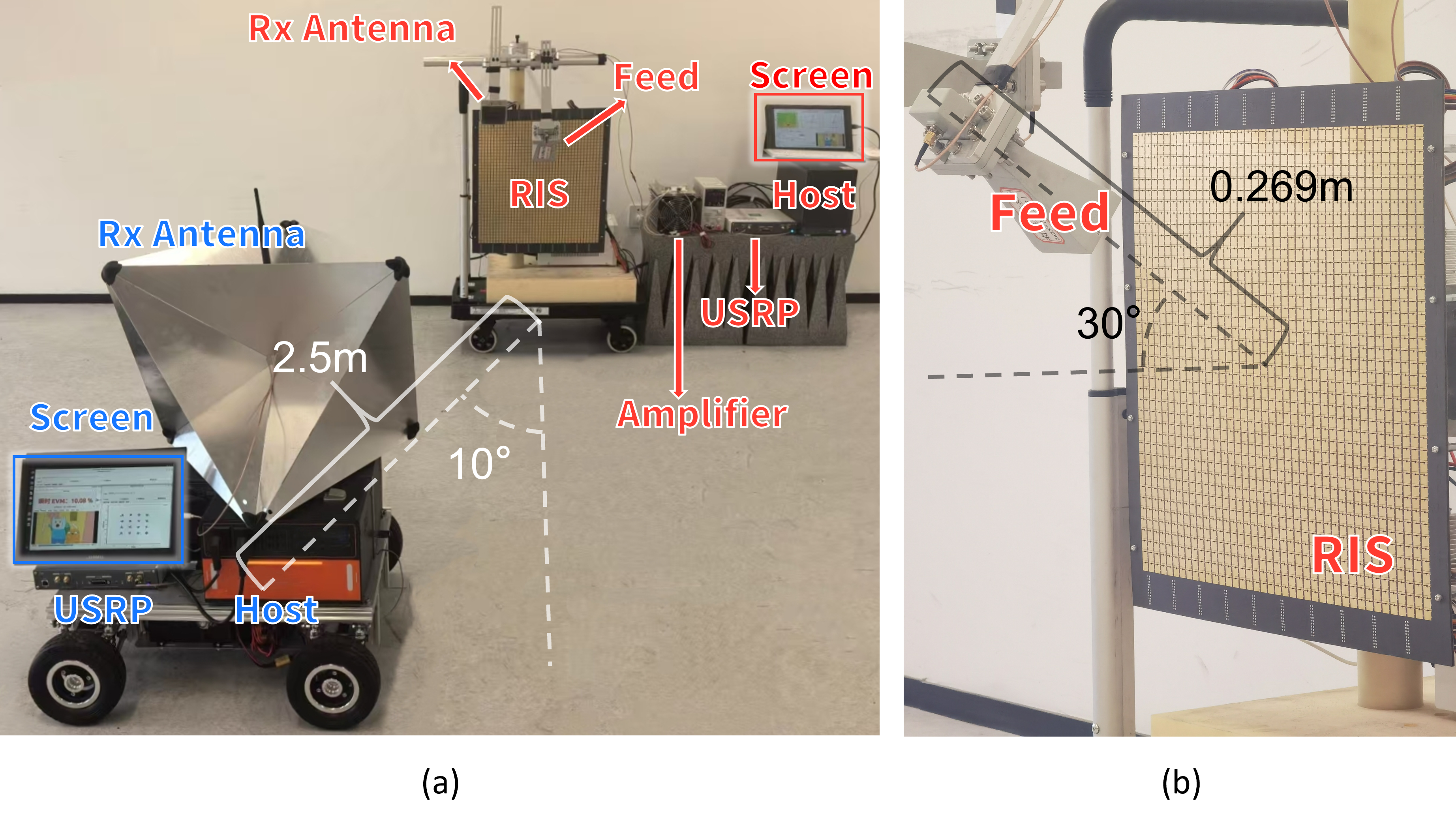}
    \vspace{-0.5cm}
	\caption{(a) Hardware modules of the implemented RIS-based ISAC-empowered vehicular network prototype and experimental layout; (b) Close-up view of the RIS-based large-scale antenna array.}
	\label{real}
    \vspace{-0.5cm}
\end{figure}

\subsection{Implementation of RIS-based Energy-efficient Large-scale Antenna Array}
The RIS integrates a large number of compact and cost-effective sub-wavelength reflecting elements into a planar aperture, enabling high beamforming gain with low hardware cost and power consumption. In our prototype, we adopt a custom-designed RIS consisting of $40\times 40$ reflecting elements operating at a center frequency of $5.5$~GHz, and the power consumption of the fabricated RIS is 6.8 W. As shown in Fig.~\ref{real}(b), the feed antenna is placed 0.269~m away from the RIS center with an elevation angle of 30$^\circ$, providing the incident waves for RIS reflection. Each element is equipped with diode-based tunable components, whose electromagnetic response can be switched by applying different bias voltages through the control circuit. By configuring the voltage states of all elements according to a predesigned reflection pattern, the RIS adjusts the reflection phases of the incident waves from the feed antenna, so that the reflected waves are coherently superimposed toward the desired direction. In this way, the RIS serves as a low-cost and low-power solution for the roadside V2I system.

\begin{figure}[!t]
	\centering
	\includegraphics[width=0.47\textwidth]{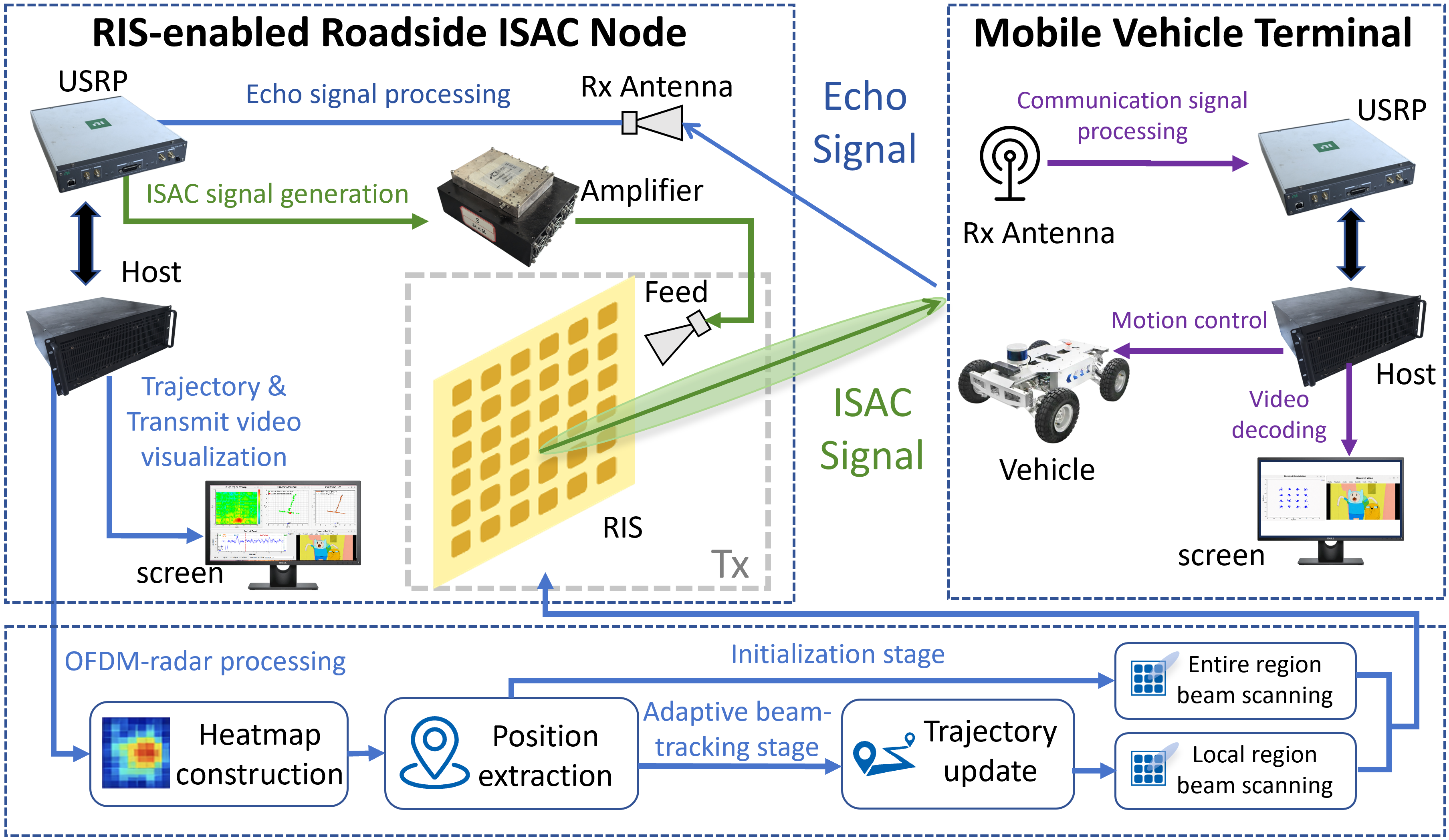}
	\caption{System structure of RIS-based ISAC-empowered vehicular network prototype.}
	\label{system_model}
\end{figure}

\begin{figure}[!t]
	\centering
	\includegraphics[width=0.48\textwidth]{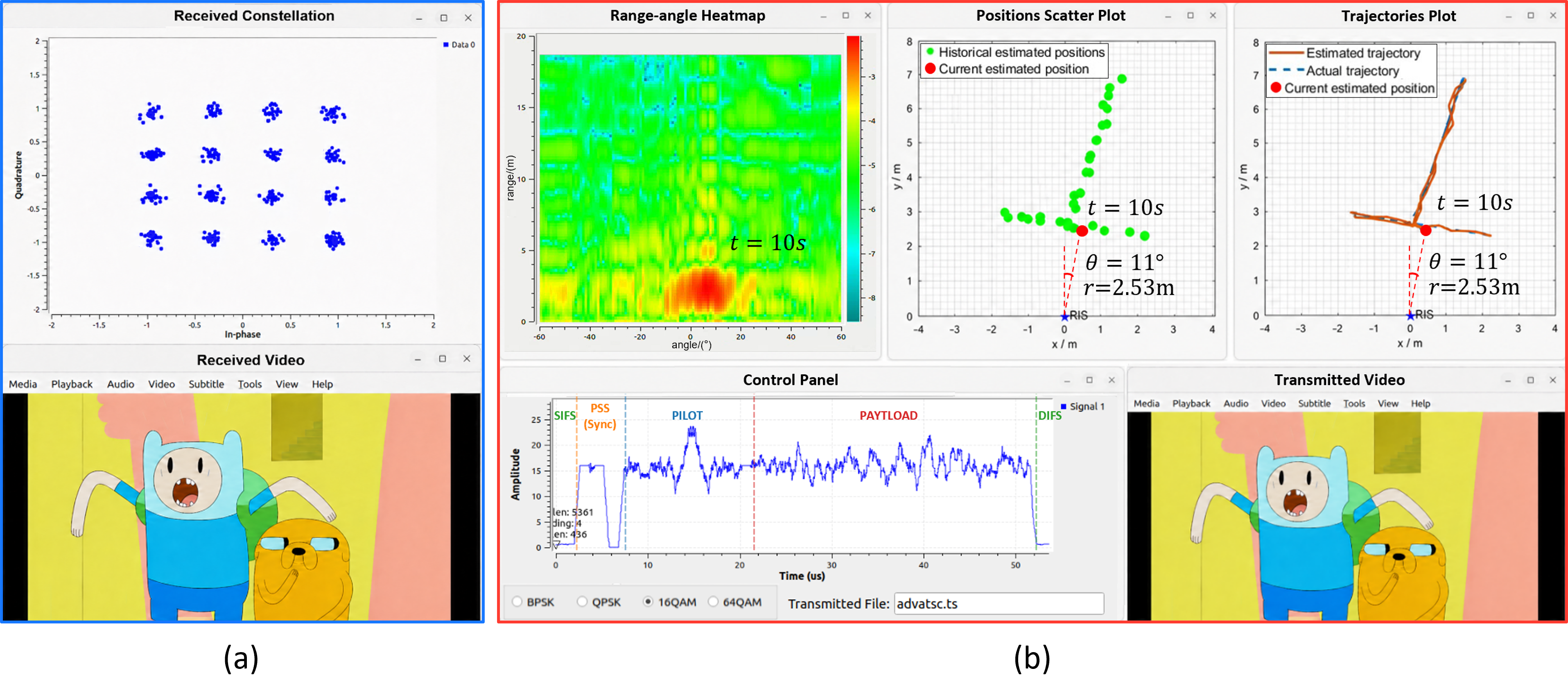}
    \vspace{-0.3cm}
	\caption{(a) Real-time interface at the vehicle terminal, which displays the demodulated constellation and the corresponding received video; (b) Real-time interface at the roadside unit. It displays the current range-angle heatmap, historical and current (the red point) estimated positions of the vehicular terminal, continuous vehicular trajectory estimated from these positions and its comparison with the actual trajectory.}
\vspace{-0.5cm}
	\label{result}
\end{figure}

\subsection{Prototyping of RIS-based ISAC-empowered Vehicular Network}

The proposed RIS-based ISAC-empowered vehicular network prototype operates at 5.5~GHz with a bandwidth of 100~MHz, and employs 16-QAM modulation for downlink data transmission. It consists of a RIS-enabled roadside node and a mobile vehicle terminal, as shown in Fig.~\ref{system_model}. Specifically, the roadside node consists of a host computer, a USRP X310, a power amplifier, a feed antenna, a RIS, and a receive (Rx) antenna. It generates the ISAC signal and processes the received echoes. The vehicle terminal is mounted on a robotic vehicle that moves along a predefined trajectory. It consists of a host computer, a USRP X310, a Rx antenna, and a robotic vehicle. It receives and demodulates the communication signal. The prototype is designed to maintain V2I communication while enabling vehicle trajectory tracking.

The overall working protocol consists of two stages, as shown at the bottom of Fig.~\ref{system_model}. In the initialization stage, the RIS-enabled roadside node transmits an OFDM-based ISAC signal and controls the RIS to perform beam scanning over the entire region. For each scanning beam, the echo signal is processed using OFDM radar processing to construct a range-angle heatmap. The CFAR algorithm is then applied to extract the initial vehicle position, based on which the roadside node configures the RIS beam toward the vehicle to establish a communication link. Then, the system enters the adaptive beam-tracking stage to periodically estimate the vehicle position over multiple tracking rounds. In each round, the RIS conducts fine beam scanning within a local angular region around the vehicle position estimated in the previous round, thereby reducing the beam-search overhead while maintaining the communication link. Compared with the initialization stage, an extended Kalman filter (EKF) is further employed to refine the extracted vehicle position and update the trajectory. The position is then used to determine the beam-scanning angular region for the next round. 

\subsection{Experimental Evaluation}

\subsubsection{Experimental Setup}
Fig.~\ref{real}(a) shows the photograph of the physical demo. The vehicle starts from a point located 3 m away from the RIS at an azimuth angle of $45^\circ$ with respect to the RIS broadside direction, then moves along a predefined trajectory, and finally stops at the ending point located 2.5 m away from the RIS at an azimuth angle of $10^\circ$. The key system parameters and experimental results are summarized in Table~\ref{tab:parameters}.
\vspace{-8pt}
\begin{table}[!h]
\centering
\caption{System parameters of the proposed prototype.}
\vspace{-3pt}
\label{tab:parameters}
\begin{tabular}{l c}
\hline
Parameter & Value \\
\hline
Carrier frequency & 5.5 GHz \\
Bandwidth & 100 MHz \\
Transmit power & 35 dBm \\
Modulation & 16-QAM \\
Number of OFDM subcarriers & 1024 \\
\hline
\end{tabular}
\end{table}
\vspace{-8pt}
\subsubsection{Experimental Results}
Figs.~\ref{result}(a) and (b) show the real-time interfaces at the vehicle terminal and the roadside node, respectively. Fig.~\ref{result}(a) displays the clear received constellation and stable video reception, indicating reliable communication during vehicle movement. Compared with the baseline without RIS, the proposed prototype reduces the error vector magnitude (EVM) from 17.35\% to 10.08\% and the block error rate (BLER) from 5.14\% to 0.73\%, indicating that the RIS effectively improves the communication quality during vehicle movement. In Fig.~\ref{result}(b), the trajectory plot shows that the estimated trajectory (orange curve) follows the actual trajectory (blue curve). The average localization error of the proposed prototype is 0.11~m, demonstrating its accurate vehicle trajectory tracking capability.

\section{Physical Venue Requirements}
This demo is preferably conducted in an open area of about 5m $\times$ 5m, while a smaller area of at least 3m $\times$ 3m is also feasible by shortening the vehicle trajectory.

\end{document}